\documentclass[prl,twocolumn,superscriptaddress,floatfix,showpacs]{revtex4}
\usepackage{amsmath,amssymb,epsfig,psfig,color,pstricks}

\renewcommand{\red}{\black}

\begin{document}

\title{$^\dagger$Explanation of the similarity of the 
photoemission spectra of SrVO$_3$ and CaVO$_3$}
\author{I.A.~Nekrasov}
\affiliation{Institute of Metal Physics, Russian Academy of Sciences-Ural Division,
620219 Yekaterinburg GSP-170, Russia}
\affiliation{Theoretical Physics III, Center for Electronic Correlations and Magnetism,
University of Augsburg, 86135 Augsburg, Germany}
\author{G.~Keller}
\affiliation{Theoretical Physics III, Center for Electronic Correlations and Magnetism,
University of Augsburg, 86135 Augsburg, Germany}
\author{D.E.~Kondakov}
\affiliation{Institute of Metal Physics, Russian Academy of Sciences-Ural Division,
620219 Yekaterinburg GSP-170, Russia}
\affiliation{Theoretical Physics III, Center for Electronic Correlations and Magnetism,
University of Augsburg, 86135 Augsburg, Germany}
\affiliation{Department of Theoretical Physics and Applied Mathematics, USTU, 620002
Yekaterinburg Mira 19, Russia}
\author{A.V.~Kozhevnikov}
\affiliation{Institute of Metal Physics, Russian Academy of Sciences-Ural Division,
620219 Yekaterinburg GSP-170, Russia}
\affiliation{Theoretical Physics III, Center for Electronic Correlations and Magnetism,
University of Augsburg, 86135 Augsburg, Germany}
\affiliation{Department of Theoretical Physics and Applied Mathematics, USTU, 620002
Yekaterinburg Mira 19, Russia}
\author{Th.~Pruschke}
\affiliation{Theoretical Physics III, Center for Electronic Correlations and Magnetism,
University of Augsburg, 86135 Augsburg, Germany}
\author{K.~Held}
\affiliation{Max Planck Institute for Solid State Research, Heisenbergstr.~1, 70569
Stuttgart, Germany}
\author{D.~Vollhardt}
\affiliation{Theoretical Physics III, Center for Electronic Correlations and Magnetism,
University of Augsburg, 86135 Augsburg, Germany}
\author{V.I.~Anisimov}
\affiliation{Institute of Metal Physics, Russian Academy of Sciences-Ural Division,
620219 Yekaterinburg GSP-170, Russia}
\affiliation{Theoretical Physics III, Center for Electronic Correlations and Magnetism,
University of Augsburg, 86135 Augsburg, Germany}

\begin{abstract}
We present parameter-free LDA+DMFT results for the many-particle
density of states of cubic SrVO$_3$ and orthorhombic CaVO$_3$.
Both systems are found to be strongly correlated metals, but {\em
not} on the verge of a metal-insulator transition. In spite of the
considerably smaller \mbox{V-O-V} bonding angle in CaVO$_{3}$ the
photoemission spectra of the two systems are very similar, their
quasiparticle parts being almost identical. This is in contrast to
earlier theoretical and experimental conclusions, but in agreement
with recent bulk-sensitive photoemission experiments.
\end{abstract}
\vspace{-0.5cm}
\maketitle

\small
{\bf{$\dagger$ Joint theoretical and experimental paper on the same subject
could be found on cond-mat/0312429.}}
\normalsize

Transition metal oxides are an ideal laboratory for the study of electronic
correlations in solids. Among these materials, cubic perovskites
have the simplest crystal structure and thus may be viewed as a starting
point for understanding the electronic properties of more complex systems.
Typically, the $3d$ states in those materials form comparatively narrow
bands of width $W\!\!\sim \!2\!-\!3\,$~eV which lead to
strong Coulomb correlations between the electrons. Particularly simple are
transition metal oxides with a 3$d^{1}$ configuration since
they do not show a complicated multiplet structure.

Intensive experimental investigations of spectral and transport properties
of strongly correlated 3$d^{1}$ transition metal oxides started with the
paper by Fujimori \textit{et al.}~\cite{Fujimori92a}. These authors observed
a pronounced lower Hubbard band in the photoemission spectra (PES) which
cannot be explained by conventional band structure theory. A number of
papers~\cite{old_experiments,Morikawa95} subsequently addressed the spectral, transport
and thermodynamic properties of the 3$d^{1}$ series Sr$_{1-x}$%
Ca$_{x}$VO$_{3}$ for various values of $x$, yielding contradictory results.
While the thermodynamic properties (Sommerfeld coefficient,
resistivity, and paramagnetic susceptibility) were found to be essentially
independent of $x$ from $x\!=\!0$ (SrVO$_{3}$) to $x\!=\!1$ (CaVO$_{3}$%
)~\cite{old_experiments}, PES and Bremsstrahlungs isochromat spectra (BIS)~\cite{Morikawa95}
 showed drastic differences.
In fact, the spectroscopic data seemed to imply that this substitution series develops
from a strongly correlated metal (SrVO$_{3}$) into, practically, an
insulator (CaVO$_{3}$), i.e., that for $x\!\rightarrow \!1$ Sr$_{1-x}$Ca$_{x}$VO$%
_{3}$ is on the verge of a Mott-Hubbard  transition.

An experimental answer to this puzzle was very recently provided by
bulk-sensitive PES obtained by Maiti~\textit{et al.}~\cite{Maiti01}, and
especially by high-resolution bulk-sensitive PES measured by
Sekiyama~\textit{et al.}~\cite{Sekiyama02}. In the latter work it was shown
that (i) the technique of preparing the sample surface (which should preferably
be done by fracturing) is very important, and that (ii) the energy of the X-ray
beam should be large enough to increase the photoelectron escape depth to
achieve bulk-sensitivity; also the beam should provide a high instrumental
resolution (about 100~meV in  \cite{Sekiyama02}). With these experimental
improvements the PES of SrVO$_{3}$ and CaVO$_{3}$ were found to be almost
identical~\cite{Maiti01,Sekiyama02}, implying consistency of spectroscopic and
thermodynamic results at last. This is also in accord with earlier 1s
x-ray absorption spectra (XAS) by Inoue {\em et al.} \cite{Inoue94} which
differ only slightly  above the Fermi energy, in contrast to the BIS data
\cite{Morikawa95}.

The main effect of substituting  Sr ions by the isovalent, but smaller, Ca ions
is to decrease the V-O-V angle from $\theta \!=\! 180^{\circ }$ in
SrVO$_{3}$~\cite{Rey90} to $\theta \!\approx\! 162^{\circ }$~\cite{InouePC} in
the orthorhombically distorted $Pbnm$ structure of
CaVO$_{3}$~\cite{Chamberland71}. This bond bending results in a decrease of the
one-particle bandwidth $W$ and thus in an increase of the ratio $U/W$ as one
moves from SrVO$_{3}$ to CaVO$_{3}$. Within a one-band Hubbard model, neglecting
the orbital structure of the V $3d$, Rozenberg~\textit{et al.}~\cite{Rozenberg96}
were able to fit the more surface-sensitive
spectroscopic data for Sr$_{1-x}$Ca$_x$VO$_{3}$  and, with other parameters, the
bulk-sensitive  PES \cite{Maiti01}.

In this Letter we present results of a parameter-free, comparative
study of SrVO$_{3}$ and CaVO$_{3}$ using the recently developed
LDA+DMFT computational 
scheme~\cite{LDA_DMFT,LDADMFTTMO1,LDADMFTTMO2,Lichtenstein01,Savrasov01,Held01},
which merges the local density approximation (LDA) with a modern many-body
technique, the dynamical mean-field theory
(DMFT)~\cite{DMFT}. In this realistic approach
the orbital degrees of freedom of the three partially filled
t$_{2g}$ bands are explicitly taken into
account~\cite{Fujimori92a}, resulting in a spectral weight distribution distinctly different
 from that of a one-band  model.

\begin{figure}[tb]
\centering \epsfig{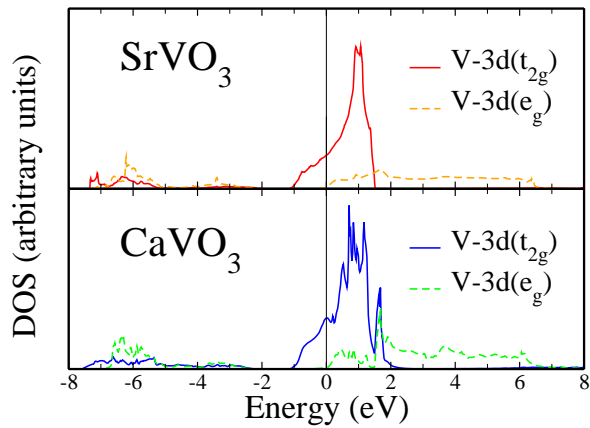}
\caption{Partial DOS of the V-3$d$ t$_{2g}$ band (solid line) and e$_{g}$
band (dashed line) for SrVO$_{3}$ and CaVO$_{3}$ calculated by the
LDA-LMTO method. The Fermi level corresponds to zero energy. \protect\vspace{%
-0.3cm}}
\label{fig_LDA_DOSes}
\end{figure}

We first discuss the electronic and crystal structure of SrVO$_{3}$ and CaVO$%
_{3}$. The valence states of these two systems consist of completely
occupied oxygen 2$p$ bands and partially occupied V-3$d$ bands. There is one
electron occupying the $d$-states per V ion ($d^{1}$ configuration).
Fig.~\ref{fig_LDA_DOSes} shows the LDA density of states (DOS) for both compounds which we
obtained via the \textsc{tblmto47} code of Andersen and coworkers~\cite%
{LMTO,comment}, using the crystal structure data of Refs.~\cite{Rey90,InouePC}.

In both compounds the V ions are in a octahedral oxygen coordination. The
octahedral crystal field splits the V-3$d$ states into three degenerate t$%
_{2g}$ and two degenerate e$_{g}$ states. While a hybridization between
these t$_{2g}$ and e$_{g}$ states is forbidden for the cubic SrVO$_{3}$
crystal, the distorted orthorhombic structure of CaVO$_{3}$ allows them to
mix. In Fig.~\ref{fig_LDA_DOSes} one can see some contributions of the t$%
_{2g}$ and e$_{g}$ subbands in the energy range from -8~eV to -2~eV due to the
hybridization with O-2$p$ states. They amount to 12\% and 15\% of all t$%
_{2g} $ states in SrVO$_{3}$ and CaVO$_{3}$, respectively. The
t$_{2g}$ states are dominant in the vicinity of the Fermi energy,
the center of gravity of the e$_{g}$ band lying above the upper
band edge of the t$_{2g}$ band. Therefore the low energy physics
($<\!1.5$~eV) is governed by the t$_{2g}$ bands which can be
considered sufficiently separated from the e$_{g}$ states in both
compounds.

\begin{figure}[tb]
\centering \epsfig{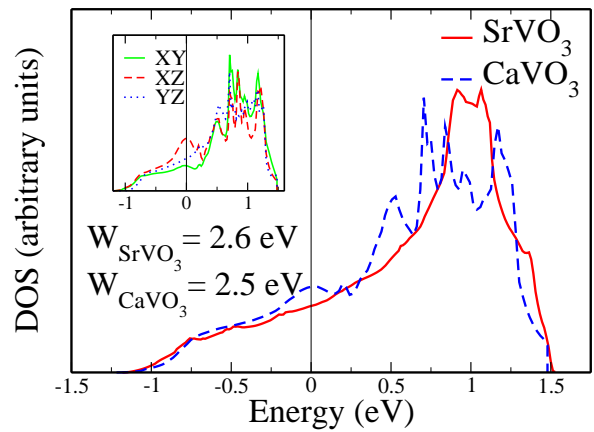}
\caption{Comparison of the LDA DOS of the V-3d t$_{2g}$ band for SrVO$_{3}$
(solid line) and CaVO$_{3}$ (dashed line). In CaVO$_{3}$ the degeneracy of
the three $t_{2g}$ bands is lifted (see inset). The bandwidth $W_{\mathrm{%
CaVO_{3}}}=2.5$eV is only 4\% smaller than $W_{\mathrm{SrVO_{3}}}=2.6~$eV.
\protect\vspace{-0.3cm}}
\label{fig_t2g}
\end{figure}

We concentrate on this energy range in Fig.~\ref{fig_t2g}, where we
compare the LDA t$_{2g}$ DOS of SrVO$_{3}$ and CaVO$_{3}$. Most importantly,
the one-electron t$_{2g}$ bandwidth  of CaVO$_{3}$, defined as the energy
interval where the DOS in Fig.~\ref{fig_t2g} is non-zero, is found to be only
4\% smaller than that of SrVO$_{3}$ ($W_{\mathrm{CaVO_{3}}}=2.5$~eV,
$W_{\mathrm{SrVO_{3}}}=2.6$~eV). One would have expected that the strong
lattice distortion with a decrease of the V-O-V bond angle from 180$^{\circ }$
to 162$^{\circ }$ affects the t$_{2g}$ bandwidth much more strongly. Such a
larger effect indeed occurs in the  e$_{g}$ bands whose bandwidth  is reduced
by 10\%. To physically understand the smallness of the  narrowing  of the 
t$_{2g}$ bands we have calculated  the effective t$_{2g}$-t$_{2g}$ and
e$_{g}$-e$_{g}$ hopping parameters. The predominant contribution to the
e$_{g}$-e$_{g}$ hopping is  through a $d\!\!-\!\!p\!\!-\!\!d$ hybridization,
which is considerably decreasing with the lattice distortion. This is also the
case for the $t_{2g}$ orbitals. However, for the  $t_{2g}$ orbitals the  direct
$d\!\!-\!\!d$ hybridization is also important. This hybridization
{\em increases} with the distortion since the  $t_{2g}$ orbital lobes point more
directly towards each other
in the distorted structure. Altogether, the
competition between  i) decreasing $d\!\!-\!\!p\!\!-\!\!d$ hybridization and 
ii) increasing  $d\!\!-\!\!d$ hybridization results in a very small change of
the t$_{2g}$ bandwidth. This explains why previous suggestions of strongly
different bandwidths are untenable.

As is well-known, LDA does not treat the effects of strong local Coulomb
correlations adequately. To overcome this drawback, we  use LDA+DMFT as a
non-perturbative approach to study strongly correlated systems~\cite%
{LDA_DMFT,Held01}. It combines the strength of the LDA in describing weakly
correlated electrons in the $s$- and $p$-orbitals, with the DMFT treatment
of the dynamics due to local Coulomb interactions. In the present paper we
will discuss the relevant parts of the LDA+DMFT approach only briefly,
refering the reader to Ref.~\cite{Held01} for details.

Given a specific material, it is possible to extract from the LDA band
structure a one-particle Hamiltonian $\hat{H}_{\mathrm{LDA}}^{0}$ where the
averaged Coulomb interaction is subtracted to avoid double counting~\cite%
{LDA_DMFT}. Supplementing $\hat{H}_{\mathrm{LDA}}^{0}$ with the local
Coulomb interactions between the electrons one arrives at a
material-specific Hamiltonian which includes correlations:%
\begin{eqnarray}
\hat{H}=\hat{H}_{\mathrm{LDA}}^{0} &+&U\sum_{m}\sum_{i}\hat{n}_{im\uparrow }%
\hat{n}_{im\downarrow }  \label{H} \\
&+&\;\sum_{i}\sum_{m\neq m^{\prime }}\sum_{\sigma \sigma ^{\prime
}}\;(U^{\prime }-\delta _{\sigma \sigma ^{\prime }}J)\;\hat{n}_{im\sigma }%
\hat{n}_{im^{\prime }\sigma ^{\prime }}.  \notag
\end{eqnarray}%
In the present case, the index $i$ enumerates the V sites, $m$ denotes the
individual t$_{2g}$ orbitals, and $\sigma $ is the spin. Because of the nearly
cubic symmetry of CaVO$_{3}$, one can simplify the calculation and use only
the one-particle LDA DOS $N^{0}(\epsilon )$ of Fig.~\ref{fig_t2g} instead
of the full Hamiltonian $\hat{H}_{\mathrm{LDA}}^{0}$~\cite{Held01}. In Eq.~(\ref{H}%
), the local intra-orbital Coulomb repulsion $U$, the
inter-orbital repulsion $U^{\prime}$, and the exchange
interaction $J$ are explicitly taken into account. We calculated
these interaction strengths by means of the constrained LDA
method~\cite{Gunnarsson89} for SrVO$_3$, allowing the e$_{g}$
states to participate in screening~\cite{Solovyev96}. The
resulting value of the averaged Coulomb interaction is
$\bar{U}=3.55$~eV ($\bar{U}=U^{\prime }$ for t${_{2g}}$
orbitals~\cite{Held01,Zoelfl00}) and $J=1.0$~eV. The intra-orbital
Coulomb repulsion $U$ is then fixed by rotational invariance to
$U=U^{\prime }+2J = 5.55$~eV.
We did not calculate $\bar{U}$ for CaVO$_{3}$ because the standard procedure
to calculate the  Coulomb interaction parameter between two t${_{2g}}$
electrons which are screened by e$_{g}$ states is not applicable 
for the distorted crystal structure where the e$_{g}$ and t$_{2g}$ orbitals
are not separated by symmetry. On the other hand, 
it is well-known that the change of the {\em local}
Coulomb interaction is  typically much smaller than the change
in the DOS, which  was found to depend only very weakly on the bond angle. 
That means that  $\bar{U}$ for CaVO$_{3}$ should be nearly
the same as for SrVO$_{3}$. Therefore we used
$\bar{U}=3.55$~eV and $J=1.0$~eV for both SrVO$_{3}$ and
CaVO$_{3}$. This is also in agreement with previous calculations
of vanadium systems~\cite{Solovyev96} and experiments
\cite{Makino98}.
\begin{figure}[tb]
\centering \epsfig{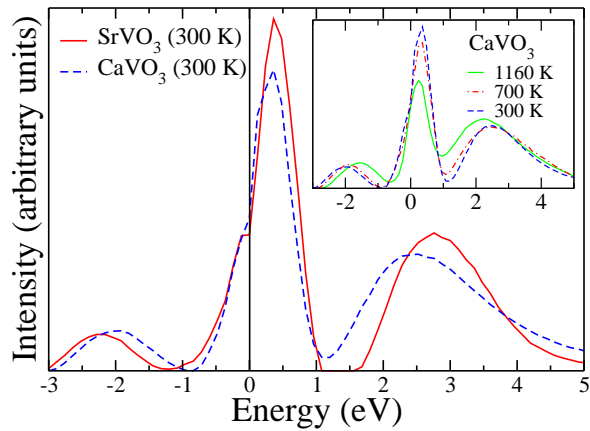}
\caption{LDA+DMFT(QMC) spectrum of SrVO$_{3}$ (solid line) and CaVO$_{3}$
(dashed line) calculated at T=300 K (inset: effect of temperature in the
case of CaVO$_{3}$).\protect\vspace{%
-0.3cm}}
\label{fig_ldadmft}
\end{figure}
\begin{figure}[tb]
\centering \epsfig{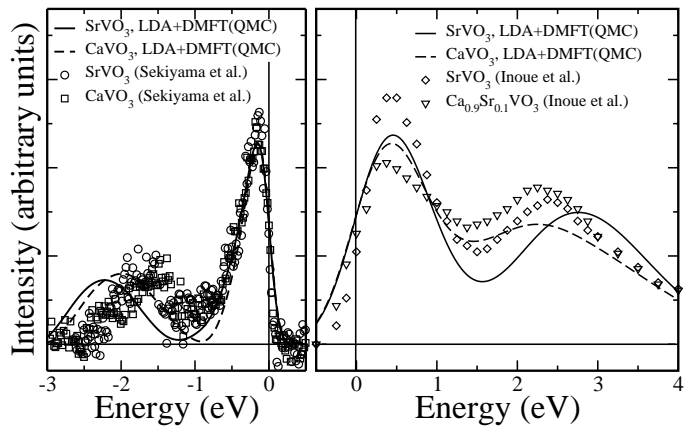}
\caption{
{\red
Comparison of the calculated, parameter-free LDA+DMFT(QMC) spectra of
SrVO$_{3}$ (solid line) and CaVO$_{3}$ (dashed line) with bulk-sensitive
high-resolution PES (SrVO$_{3}$: circles; CaVO$_{3}$: rectangles)
~\protect\cite{Sekiyama02} (left panel) and 1s XAS spectra (SrVO$_{3}$:
diamonds; Ca$_{0.9}$Sr$_{0.1}$VO$_{3}$: triangles)~\protect\cite{Inoue94}
(right panel). Horizontal line: experimental subtraction of the background
intensity.}
\protect\vspace{-0.3cm}
\label{fig_XPS}}
\end{figure}
The Hamiltonian (\ref{H}) is solved within the DMFT using standard
quantum Monte-Carlo (QMC) techniques to solve the self-consistency
equations~\cite{QMC}. From the imaginary time QMC Green function
we calculate the physical (real frequency) spectral function with
the maximum entropy method. The resulting LDA+DMFT(QMC)
spectra for SrVO$_{3}$ and CaVO$_{3}$ in Fig.~\ref{fig_ldadmft}
show genuine correlation effects, i.e., the formation of lower
Hubbard bands at about $-1.5$~eV and upper Hubbard bands at about
$2.5$~eV, with well pronounced quasiparticle peaks at the Fermi
energy. Therefore both SrVO$_{3}$ and CaVO$_{3}$ are strongly
correlated metals. The 4\% difference in the LDA bandwidth between
SrVO$_{3}$ and CaVO$_{3}$ is only reflected in some additional
transfer of spectral weight from the quasiparticle peak to the
Hubbard bands, and minor differences in the positions of the
Hubbard bands. Clearly, the two systems are not on the verge of a
Mott-Hubbard metal-insulator transition. The many-particle DOS of
the two systems shown in Fig.~\ref{fig_ldadmft} are seen to be
quite similar but not identical. In fact, SrVO$_{3}$ is slightly
less correlated than CaVO$_{3}$, in accord with their different
LDA bandwidth. The inset of Fig.~\ref{fig_ldadmft} shows that the
effect of temperature on the spectrum is small for $T \lesssim
700$~K. Due to the high photon energy the PES transition matrix
elements will not significantly affect the distribution of
spectral weight in the energy range considered here.

In the left panel of Fig.~\ref{fig_XPS} we compare our LDA+DMFT spectra
(300K), which were multiplied with the Fermi function at the experimental
temperature (20$\,$K) and Gauss broadened with the experimental resolution of
$0.1\,$eV \cite{Sekiyama02}, to the experimental PES data obtained by
subtracting estimated surface and oxygen contributions. The quasiparticle peaks
in theory and experiment are seen to be in very good agreement. In particular,
their height and width are almost identical for both SrVO$_{3}$ and CaVO$_{3}$.
The difference in the positions of the lower Hubbard bands may be partly due to
(i) the subtraction of the (estimated) oxygen contribution which might also
remove some $3d$ spectral weight below $-2$~eV, and (ii) uncertainties in the
{\em{ab-initio}} calculation of $\bar{U}$. In the right panel of 
Fig.~\ref{fig_XPS} we compare to the XAS data~\cite{Inoue94}. We consider
core-hole life time effects by Lorentz broadening the spectrum with
0.2~eV~\cite{Krause79}, multiplying with the inverse Fermi function (80K), and
then Gauss broadening with the experimental resolution of
$0.36\,$eV~\cite{Inoue03}. Again, the overall agreement of the weights and
positions of the quasiparticle and upper $t_{2g}$ Hubbard band is good,
including the tendencies when going from  SrVO$_{3}$ to CaVO$_{3}$
(Ca$_{0.9}$Sr$_{0.1}$VO$_{3}$ in the experiment ). For  CaVO$_{3}$ the weight
of the quasiparticle peak is somewhat lower than in the experiment. In contrast
to the one-band Hubbard model calculation, our material specific results
reproduce the strong asymmetry around the Fermi energy w.r.t. weights and
bandwidths. Our results also give a different interpretation of the XAS than
in~\cite{Inoue94} where the maximum at about $2.5\,$eV was attributed to a
$e_g$ band and not to the $t_{2g}$ upper Hubbard band \cite{eg}. The slight
differences in the quasiparticle peaks (see Fig.~\ref{fig_ldadmft}) lead to
different effective masses, namely $m^*/m_0\!=\!2.1$ for SrVO$_{3}$ and
$m^*/m_0\!=\!2.4$ for CaVO$_{3}$. These theoretical values agree with $m^{\ast
}/m_{0}\!=\!2-3$ for SrVO$_{3}$ and CaVO$_{3}$ as obtained from de Haas-van
Alphen experiments and thermodynamics \cite{old_experiments,Inoue02}. We note
that the effective mass of CaVO$_{3}$ obtained from optical experiments is
somewhat larger, i.e., $m^{\ast }/m_{0}\!=\!3.9$ \cite{Makino98}.

In summary, we investigated the spectral properties of the correlated 3$d^{1}$
systems SrVO$_{3}$ and CaVO$_{3}$ within the LDA+DMFT(QMC) approach. 
Constrained LDA was used to determine the average Coulomb interaction as
$\bar{U}\!=\!3.55$~eV and the exchange coupling as $J\!=\!1.0$~eV. With this
input we calculated the spectra of the two systems in a parameter-free way. Both
systems are found to be strongly correlated metals, with a transfer of most of
the spectral weight from the quasiparticle peak to the incoherent upper and
lower Hubbard bands. Although the calculated DMFT spectra of SrVO$_{3}$ and
CaVO$_{3}$ are slightly different above the Fermi energy the spectra below the
Fermi energy are the resulting PES are very similiar, the quasiparticle parts
being almost identical. Our calculated spectra agree very well with recent
bulk-sensitive high-resolution PES \cite{Sekiyama02} and XAS  \cite{Inoue94},
i.e., with the experimental spectrum below {\em and} above the Fermi energy.
Both compounds are similarly strongly correlated metals; CaVO$_{3}$  is not on
the verge of a Mott-Hubbard transition.

Our results are in striking contrast to previous theories and the  widespread
expectation that the strong lattice distortion leads to   a strong narrowing of
the   CaVO$_{3}$ bandwidth and, hence, much stronger correlation effects in
CaVO$_{3}$. While the  e$_g$ bands indeed narrow  considerably, the competition
between decreasing $d\!\!-\!\!p\!\!-\!\!d$  and increasing $d\!\!-\!\!d$
hybridization leads to a rather insignificant narrowing of the t$_{2g}$ bands at
the Fermi energy. This explains why  CaVO$_{3}$  and SrVO$_{3}$ are so similar.
With our theoretical results confirming the  new  PES and XAS experiments, we
conclude that  the insulating-like behavior observed in earlier PES and BIS
experiments on CaVO$_{3}$ must be attributed to surface
effects~\cite{Liebsch02}.

\begin{acknowledgments}
\red{This research was supported in part by the Deutsche
Forschungsgemeinschaft through SFB 484 and the
Emmy-Noether program, the Russian Foundation for Basic
Research through RFFI-01-02-17063 and RFFI-03-02-06126,
the Ural Branch of the Russian Academy of Sciences for
Young Scientists, Grant of the President of Russia MK-95.2003.02,
and the National Science Foundation
under Grant No. PHY99-07949. We thank A. Sandvik for
making available his maximum entropy code and acknowledge valuable
discussions with I.~H.~Inoue, S.~Suga, A.~Fujimori, R.~Claessen,
and M.~Mayr.}
\end{acknowledgments}


\begin{thebibliography}{99}
\bibitem{eg}From the centers of gravity of the LDA DOS
(Fig.~\ref{fig_LDA_DOSes}),  we would expect the correlated $e_g$ states to
form an upper Hubbard band with  center of gravity at roughly  $2.5\,$eV above
the upper Hubbard band of the  $t_{2g}$ states, with the  $e_g$ Hubbard band
being very flat.
\end{thebibliography}
\end{document}